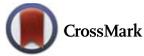

PAPER

# Dephasing-tolerant quantum sensing for transverse magnetic fields with spin qudits

Matteo Mezzadri[1,2,3,4], Luca Lepori[1,2,3,4], Alessandro Chiesa[1,2,3] and Stefano Carretta[1,2,3,*]

[1] Dipartimento di Scienze Matematiche, Fisiche e Informatiche, Università di Parma, I-43124 Parma, Italy
[2] Gruppo Collegato di Parma, INFN-Sezione Milano-Bicocca, I-43124 Parma, Italy
[3] UdR Parma, INSTM, I-43124 Parma, Italy
[4] These authors contributed equally to the present work.
[*] Author to whom any correspondence should be addressed.

E-mail: stefano.carretta@unipr.it

## Abstract

We propose a dephasing-tolerant protocol for quantum sensing of transverse magnetic fields which exploits spin qudit sensors with embedded fault-tolerant (FT) quantum error correction. By exploiting longitudinal drives, the transverse field induces logical Rabi oscillations between encoded states, whose frequency is linear in the transverse field to be probed. Numerical simulations show that the present FT protocol enables the detection of very small fields, orders of magnitudes below the limit imposed by the coherence time.

## 1. Introduction

Quantum sensing exploits intrinsically quantum properties such as superposition and entanglement to estimate unknown quantities with sensitivities impossible for classical meters [1–4], with possible applications in pure and applied research and in industries [5, 6]. Moreover, thanks to their nanoscale dimensions, quantum sensors such as atoms and molecules often permit a much higher spatial resolution.

Magnetometry is a natural playground for quantum sensing. In particular, longitudinal DC magnetic fields $B_z$ can be probed by Ramsey interferometry [2, 7], which detects the phase ($\propto B_z$) accumulated by a superposition of the two states of a spin 1/2 during free evolution. Conversely, measuring tiny transverse fields $B_x$ is much more challenging, since the effect on the accumulated phase is quadratic (not linear) in $B_x$ [8] and thus requires long interrogation times, which make the system vulnerable to decoherence. Therefore, by setting an upper bound for the interrogation time, decoherence severely limits the achievable sensitivities [2].

To cope with this issue, a frequency up-conversion of $B_x$ was implemented on NV centers [8], by mixing $B_x$ with an oscillating transverse bias field and then realizing its lock-in detection. These schemes can be combined with dynamical decoupling [8–10], thus removing quasistatic noise and improving the sensitivity of magnetometers below the limit forced by $T_2^*$. However, dynamical decoupling does not remove high-frequency noise and can result in a simultaneous suppression of any low-frequency component, including both noise and any DC signal [11].

In recent years, a more powerful alternative has emerged, potentially bringing the sensor sensitivity even well below the limit caused by the true coherence time $T_2$. The idea is to use ad hoc quantum error correction (QEC) techniques to increase the coherence of the system during a sensing protocol [11–16]. However, this idea carries on some crucial issues which must be taken into account.

In particular, straightforward application of the correction procedure can introduce a bias [15] in the measured quantity. Furthermore, a QEC scheme aimed only at the errors limiting the sensitivity of the sensing protocol is needed. In fact, a QEC procedure that corrects for a too broad class of errors could end up correcting also the evolution induced by the quantity to be measured [14, 17]. An effective solution to this problem is to design a protocol such that the signal induces a logical operation on the encoded system. This ensures the signal to be clearly distinguished from the errors corrected by the code and translates into the fault-tolerant (FT) implementation of a continuous logical operator. FT quantum computing (QC) ensures that errors are suppressed at any step of the computation, and in particular during the sensing protocol. The realization of a continuous logical operator on a multi-qubit stabilizer code was investigated in [16], but a FT implementation was not found, thus strongly limiting the error correction efficiency.

Here we propose a protocol for quantum sensing of transverse magnetic fields *not limited by* $T_2$, in which the sensor is a multi-level *spin qudit* working as a logical qubit (LQ) with embedded FT quantum error-correction [18]. In the protocol, the transverse $B_x$ induces a FT *logical Rabi* oscillation between error-protected states, by exploiting a multi-frequency longitudinal drive [19, 20]. Thus, QEC does not correct the evolution induced by $B_x$, allowing for an effective detection. Remarkably, the corresponding logical Rabi frequency is linear in $B_x$. Moreover, a crucial aspect of our FT procedure is that all the operations (both error correction and coherent rotations involved in the sensing protocol) result in the same unitaries within the logical and error subspaces, thus removing any bias onto the measured field [15].

A prototypical realization of the spin qudit sensor is represented by molecular nanomagnets (MNMs) [21–24]. Indeed, their intrinsic multi-level structure, long coherence times, addressability of individual pairs of levels by electromagnetic drives and unparalleled degree of chemical tunability recently led to propose them as qubits with embedded QEC [24–36]. In particular, they can display the proper connectivity between energy eigenstates which enables to implement QEC also during logical gates, i.e. to achieve FT computing [18]. Furthermore, they were investigated in the context of quantum estimation theory [37] and, recently, they were employed as ensemble sensors of AC-fields [38].

By combining these ingredients (i.e. quantum sensing with a longitudinal drive, a scheme for FT computing and the identification of a suitable platform), we introduce a dephasing-tolerant protocol able to detect tiny transverse magnetic fields with high sensitivity. We demonstrate this by thorough numerical simulations of the full protocol, explicitly including decoherence and the sequence of pulses required for sensing and error correction.

## 2. Basic idea on a spin qubit

The first ingredient of our protocol is a longitudinal driving field. To illustrate its effect [39], we start by considering a spin qubit sensor, described by the following Hamiltonian

$$H = H_0 + H_1 = \Delta s_z + g\mu_B B_x s_x. \tag{1}$$

Here $s_\alpha = \sigma_\alpha/2$ are spin 1/2 operators, $\mu_B$ is the Bohr magneton, $\hbar \equiv 1$, $g$ is the $g$-factor (assumed isotropic for simplicity) and $B_x$ is the *tiny* magnetic field we aim to measure. The energy splitting along the quantization axis $\Delta$ typically arises from the Zeeman interaction with a *large* magnetic field $B_z$, i.e. $\Delta = g\mu_B B_z$ and $B_x \ll B_z$. Hence, the eigenstates of the spin qubit are practically the same of $H_0 \propto s_z$ (the usual computational basis $|0\rangle$ and $|1\rangle$) and the qubit prepared in $|0\rangle$ does not undergo any evolution. Conversely, if prepared in a superposition $|+\rangle = (|0\rangle + |1\rangle)/\sqrt{2}$, it accumulates a relative phase between the two components which can be detected by moving back to the computational basis according to the Ramsey scheme [2]. This phase is linear in $B_z$, with only a small correction quadratic in $B_x$.

In order to make *the effect of $B_x$ on some observable linear*, we need to compensate the large energy gap $\Delta$. This can be achieved by means of a longitudinal drive of the form

$$h_1(t) = g\mu_B \cos(\omega_z t + \phi_z) B_{1z} s_z. \tag{2}$$

The effect of this term on the qubit time evolution can be understood by moving to a frame rotating at frequency $\omega_z/2\pi$ [19, 20, 40] (see appendix A) where the effective Hamiltonian $H_{\rm rf}$ becomes time independent to first order in the small parameter $g\mu_B B_{1z}/\omega_z$. In this limit, multi-photon transitions and fast rotating transverse terms can be omitted, thus obtaining

$$H_{\rm rf} \approx (\Delta - \omega_z) s_z - \frac{g^2 \mu_B^2 B_x B_{1z}}{\omega_z} s_x. \tag{3}$$

Details on the exact form of $H_{\rm rf}$ and on its derivation are reported in appendix A. From Hamiltonian (3) it is clear that on resonance ($\omega_z = \Delta$) the energy splitting $\Delta$ is removed in the rotating frame and the system undergoes Rabi oscillations between $|0\rangle$ and $|1\rangle$ at angular frequency

$$\Omega_R = \frac{g^2 \mu_B^2 B_{1z}}{\omega_z} B_x. \tag{4}$$

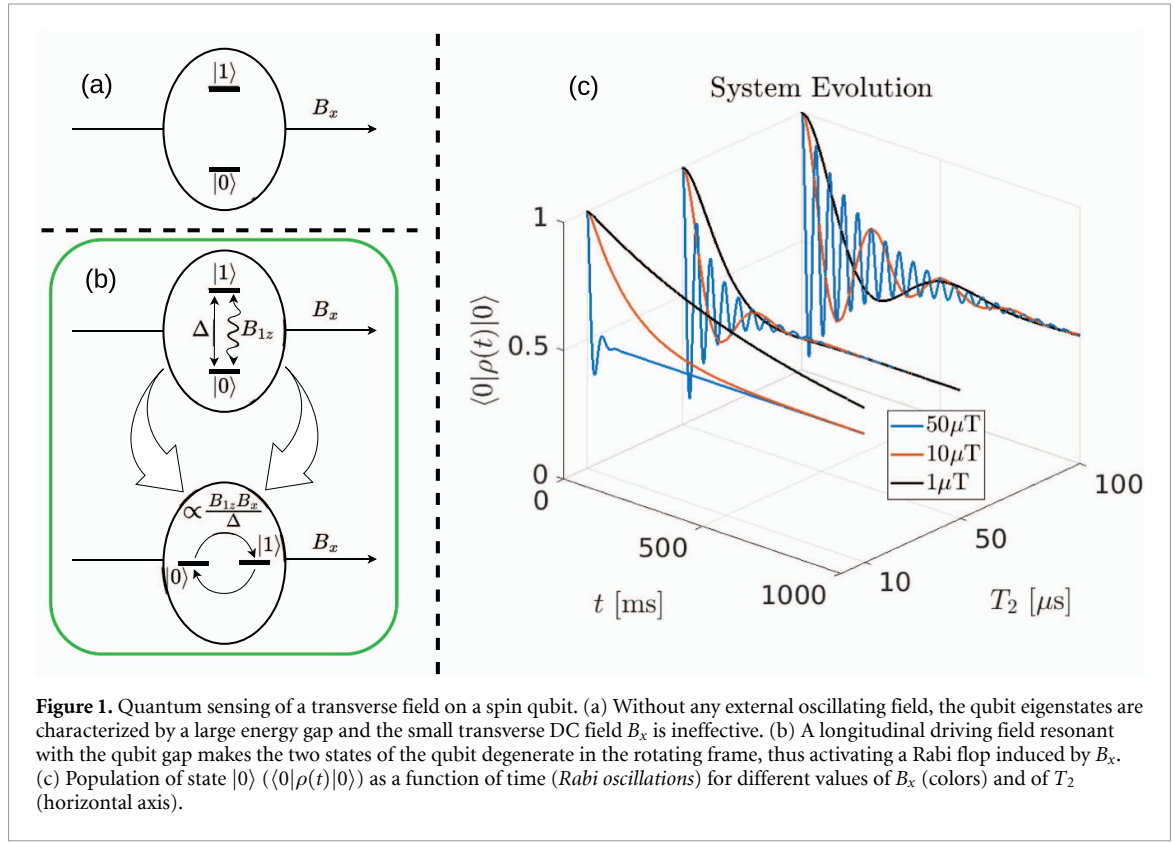

**Figure 1.** Quantum sensing of a transverse field on a spin qubit. (a) Without any external oscillating field, the qubit eigenstates are characterized by a large energy gap and the small transverse DC field $B_x$ is ineffective. (b) A longitudinal driving field resonant with the qubit gap makes the two states of the qubit degenerate in the rotating frame, thus activating a Rabi flop induced by $B_x$. (c) Population of state $|0\rangle$ ($\langle 0|\rho(t)|0\rangle$) as a function of time (*Rabi oscillations*) for different values of $B_x$ (colors) and of $T_2$ (horizontal axis).

It is worth noting that $\Omega_R$ is proportional to $B_x$, and hence the time evolution of the system is completely ruled by $B_x$. This is substantially different from the scheme proposed in [8], where the transverse field induces only a small shift in the frequency of the observed oscillations and hence a good estimate of $B_x$ requires to observe a large number of oscillations in the populations of the system eigenstates. With the proposed protocol, we can directly relate the period of an oscillation to the transverse field to be measured, thus limiting to just a single oscillation of the two states without moving to the frequency domain. In the absence of decoherence, equation (4), or the analogous one for the Rabi period $\tau(B_x) = 2\pi/\Omega_R$, directly yields $B_x$.

The idea is sketched in figures 1(a) and (b): in the absence of the external pulse $h_1(t)$ (a), the qubit shows a large energy gap between the eigenstates $|0\rangle$ and $|1\rangle$. Therefore, the small $B_x$ does not induce any evolution on the system prepared, e.g. in $|0\rangle$. As soon as the longitudinal drive is turned on (b), the qubit energy levels are made effectively degenerate and hence the system displays oscillations between $|0\rangle$ and $|1\rangle$ with angular frequency $\Omega_R$. Compared to a degenerate qubit with $\Delta = 0$, $\Omega_R$ is re-scaled by a factor $\alpha = g\mu_B B_{1z}/\omega_z$. In resonance and in presence of a longitudinal DC field $\omega_z = \Delta = g\mu_B B_z$ and hence $\alpha = b_{1z}/B_z \sim 10^{-2}$. As a result, Rabi nutations are rather slow and can be significantly damped by the effect of decoherence.

To investigate this issue, we perform numerical simulations of the qubit time evolution due to the Hamiltonian $H + h_1(t)$, including pure dephasing in Lindblad form. The corresponding master equation for the system density matrix $\rho$ is given by

$$\dot{\rho} = -i\left[H + h_1(t), \rho\right] + \frac{1}{T_2}\left(2s_z\rho s_z - s_z^2\rho - \rho s_z^2\right), \tag{5}$$

where the first term represents the coherent evolution, while the second models pure dephasing with rate $1/T_2$. We also include a readout time $t_m$ at the end of the time evolution, during which the system is still subject to pure dephasing. The resulting population on state $|0\rangle$, $\langle 0|\rho|0\rangle$, obtained by solving equation (5), is reported in figure 1(c) as a function of time, for different values of $T_2$ (horizontal axis) and $B_x$ (colors).

Intuitively, the minimum $B_x$ which allows us to see a complete Rabi flop decreases by increasing $T_2$: a single damped oscillation is visible for $T_2 = 10\,\mu\text{s}$ and $B_x = 50\,\mu\text{T}$, while we need $T_2 \sim 50$–$100\,\mu\text{s}$ to distinguish complete oscillations with $B_x = 10$–$1\,\mu\text{T}$.

As mentioned above, a peculiar feature of our protocol is that we retrieve information about the external field by direct observation of slow Rabi oscillations. To estimate $\tau$ we consider the time at which the populations of an initial state prepared in $|0\rangle$ and $|1\rangle$ cross. In the ideal case without dephasing, this time is exactly $\tau/4$ and corresponds to the point of maximum derivative [2]. Decoherence dampens Rabi oscillations, moving the crossing point to longer times for smaller $\Omega_R T_2 \propto B_x T_2$. The curve $B_x(\tau)$ resulting from the inversion of the monotonic function $\tau(B_x)$ corresponds to the *calibration of the sensor*. The calibration curve obtained from our simulations on a spin qubit is reported in figure 2. We note that $B_x \propto 1/\tau$ without decoherence (black line), while in presence of pure dephasing the behavior deviates from the ideal one but remains detectable even for longer $\tau$. The red curve is computed with $T_2 = 50\,\mu s$, a reasonable coherence time for molecular spin qubits [41–43]. The other parameters in the simulation are also typical of molecular spins, with $\Delta = 0.04$ meV $= 9.8$ GHz (i.e. $B_z = 0.35$ T), isotropic $g = 2$ and $B_{1z} = 10$ mT [44].

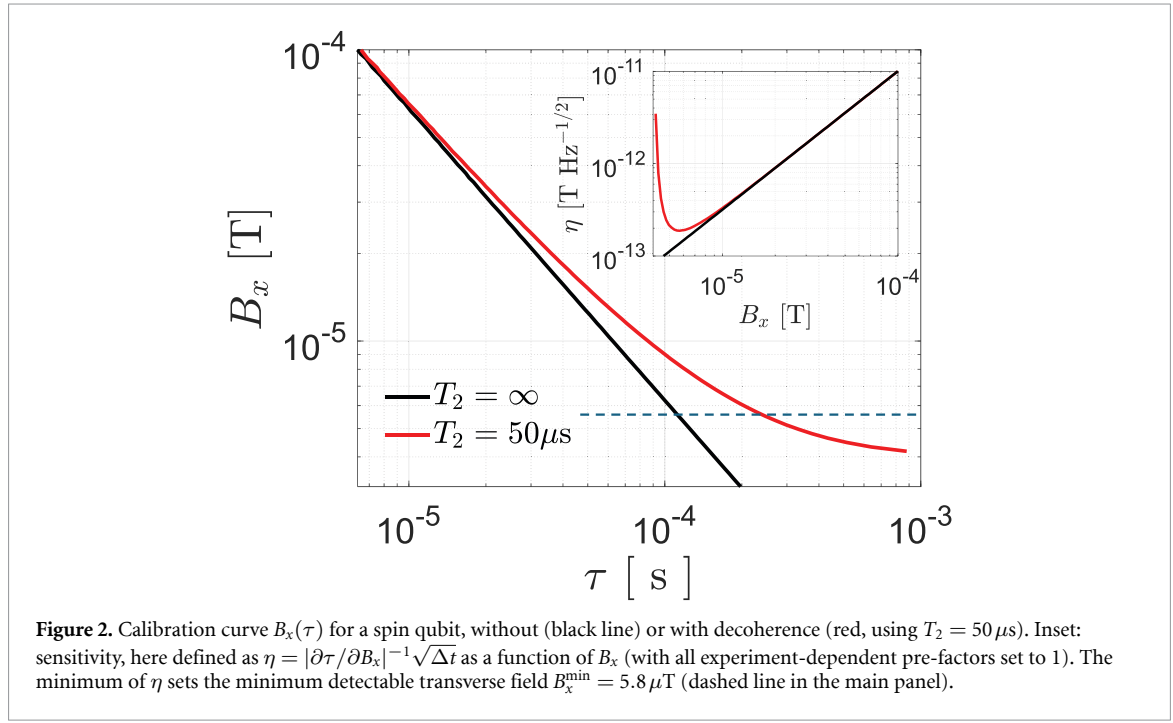

**Figure 2.** Calibration curve $B_x(\tau)$ for a spin qubit, without (black line) or with decoherence (red, using $T_2 = 50\,\mu s$). Inset: sensitivity, here defined as $\eta = |\partial\tau/\partial B_x|^{-1}\sqrt{\Delta t}$ as a function of $B_x$ (with all experiment-dependent pre-factors set to 1). The minimum of $\eta$ sets the minimum detectable transverse field $B_x^{\min} = 5.8\,\mu T$ (dashed line in the main panel).

It is now useful to recall the notion of *sensitivity* $\eta$ [2, 45] of a quantum sensor. This quantity measures the precision with which a signal is detected, thereby quantifying the quality of a sensing protocol for any field to be measured. To evaluate the quality of the protocol itself, one should ideally consider both the minimum field at which the protocol works and the precision with which this field can be measured.

In general, the sensitivity $\eta$ is the minimum detectable variation of the signal that gives unitary signal-to-noise ratio per unit integration time. This can be obtained as the sum of classical (statistical) and quantum errors gathered in $N$ cycles (each one composed by the evolution of the initially prepared quantum state and by the consequent quantum measurement) performed in the unitary interrogation time. Referred to our protocol, this leads to the following expression (see [2, 45–48] and references therein):

$$\eta(B_x, \Delta t) \propto \left|\frac{\partial\tau}{\partial B_x}\right|^{-1} \frac{1}{\sqrt{N}} \propto \left|\frac{\partial\tau}{\partial B_x}\right|^{-1} \sqrt{\Delta t}, \quad (6)$$

where $\Delta t$ is the total time for a single measurement cycle, including in our scheme both the duration of the pulses needed by the sensing protocol and the readout time $t_m$, and $N$ is the number of cycles per unit time. Therefore, apart from constant factors specifically depending on the experimental platform (such as $N$ and all other classical and quantum noise contributions like spin projection noise [2], which are not our focus here), $\eta$ is proportional to $|\frac{\partial\tau}{\partial B_x}|^{-1}$ and to $\sqrt{\Delta t}$.

Note that this definition of sensitivity allow us to get rid of details related to the specific experimental implementation and to directly focus on the main subject of our investigation, i.e. the possibility to enhance $\eta$ by suppressing decoherence via quantum error-correction. Indeed, the crucial quantity here is $|\frac{\partial\tau}{\partial B_x}|^{-1}$, which is computed explicitly from the $\tau(B_x)$ obtained from our simulations (including pure dephasing). Without decoherence (black curve in the inset of figure 2), $\partial\tau/\partial B_x \propto B_x^{-2}$, while (neglecting $t_m \ll \tau$) $\eta \propto B_x^{3/2}$. In presence of pure dephasing, $\eta$ displays a minimum in the region in which $B_x(\tau)$ deviates from a power law. We define the value of $B_x$ corresponding to this minimum as the minimum detectable transverse field, $B_x^{\min}$, due to the steep rise of $\eta(B_x)$ for smaller $B_x$. Here we obtain $B_x^{\min} \approx 5.8\,\mu T$ (dashed line in the main panel).

To summarize, although this protocol employing a longitudinal drive makes the measured Rabi frequency linear in $B_x$, both the minimum detectable transverse field and the sensitivity are strongly limited by decoherence.

## 3. Beating dephasing by qudit-embedded QEC

Hereafter, we show that dephasing can be suppressed by nesting QEC within the sensing protocol, thus effectively enhancing $T_2$ by orders of magnitude. To this aim, we resort on recently proposed schemes for qudit-based QEC on molecular spins [18, 32–35]. By embedding QEC within single objects and not into a collection of many interacting qubits, these codes avoid the resource overhead of multi-qubit approaches, while achieving an even better performance against biased noise [18, 35]. Very recently, we put forward a FT implementation of an embedded QEC code fighting the leading error on molecular spin qudits, i.e. pure dephasing [18].

Here we combine this approach with the sensing protocol, i.e. we replace the Rabi oscillation between the two levels of a physical qubit ($|0\rangle$ and $|1\rangle$) with a logical rotation between encoded logical states $|0_L\rangle$ and $|1_L\rangle$, induced by the transverse field we aim to detect.

Below we first introduce the basic principles of FT-QC and then we illustrate step-by-step how to (i) encode error-protected states into proper superpositions of eigenstates of a spin qudit; (ii) manipulate the qudit fault-tolerantly to yield logical Rabi oscillations induced by $B_x$ and (iii) correct dephasing errors during the logical evolution of the system. Being FT, our implementation does not propagate the leading errors (corrected by the code) during the *logical Rabi* rotations, thus enabling correction of these errors during the dephasing-tolerant sensing protocol.

### 3.1. Fault-tolerant Quantum Computing in a nutshell

The effect of noise on the evolution of a physical $d$-levels system can be described by the Kraus map $\rho(t) = \sum_{k=0}^{d-1} E_k(t) \rho_0 E_k^\dagger(t)$, where $E_k$ are error operators. Our Embedded QEC starts from the identification of a subset of leading errors $\{E_k, k \leqslant d/2 - 1\}$ which we aim to correct. To correct for such errors one must find proper superpositions of the system eigenstates (logical states $|0_L\rangle$ and $|1_L\rangle$), which satisfy Knill–Laflamme conditions:

$$\langle 0_L | E_k^\dagger E_j | 0_L \rangle = \langle 1_L | E_k^\dagger E_j | 1_L \rangle \tag{7a}$$

$$\langle 0_L | E_k^\dagger E_j | 1_L \rangle = 0. \tag{7b}$$

Conditions (7) guarantee that the set of errors $E_k$ can be detected and corrected. However, the actual implementation of this correction, together with logical gates between encoded states, is not trivial. In particular, the effectiveness of QEC requires that all these manipulations are realized without introducing additional errors not managed by the code. In other words, these procedures must be *error transparent* [18, 31].

To formalize this, we consider the orthogonal set of *error words* $\{|\ell, k\rangle\}$ obtained from orthogonalization of $\{E_k|0_L\rangle, E_k|1_L\rangle\}$, where $\ell = 0, 1$ labels the logical subspace while $k = 0, \ldots, d/2 - 1$ indicates the error subspace. A gate implementation is FT if the evolution within the logical subspace is independent of the error $k$ and in particular it is the same as in the no-error subspace ($k = 0$) [31]. In other words, we must ensure that operations on the system do not mix different error subspaces, generating superpositions of $|\ell, k\rangle$ and $|\ell', k'\rangle$ with $k \neq k'$. Thus, given a gate $U$ on a qubit, a proper extension of such gate for an embedded FT-QEC scheme is given by $\mathscr{U} = U \otimes \mathbb{I}_{d/2}$ [18]. We have recently shown that for embedded QEC and pure dephasing errors ($T_2$) these requirements translate into a proper connectivity between the system eigenstates, i.e. the capability to drive direct transitions between all of them via resonant pulses [18]. See details in appendix D or [18] for a more comprehensive treatment.

Along the same lines one can develop Error Transparent procedures for error detection and correction, as detailed in section 3.5.

### 3.2. Qudit system and implementation

Our goal is to find a qudit system which can support QEC by suitable code words and then let the tiny transverse field $B_x$ induce a FT *logical Rabi* oscillation between $|0, k\rangle$ and $|1, k\rangle$. Here we apply our idea to a spin $S$ qudit characterized by the Hamiltonian $H_d = H_d^{(0)} + H_d^{(1)}$, with

$$H_d^{(0)} = g\mu_B B S_z + D S_z^2 + E\left(S_x^2 - S_y^2\right) \tag{8a}$$

$$H_d^{(1)} = g\mu_B B_x S_x. \tag{8b}$$

Here $S \geqslant 3/2$ is the half-integer spin of the qudit, originating either from a single magnetic ion or from several spins giving rise to a single-molecule magnet characterized by a spin $S$ ground multiplet. The leading Hamiltonian $H_d^{(0)}$ contains, besides the Zeeman coupling with a longitudinal field $B$, also axial and rhombic zero-field splitting terms, parameterized by *D* and *E*, respectively. $H_d^{(1)}$ is the Zeeman interaction with the small $B_x$ we aim to probe.

Molecular spin systems, and in particular MNMs [24, 49] are especially suitable for our scheme. Indeed, they are typically characterized by both a clear hierarchy of error operators, where dephasing is largely dominant, and by a zero-field splitting Hamiltonian of the form (8*a*) with the presence of both *D* and *E*.

We stress that a broad class of simple MNMs satisfy the above requirements [50–52]. In particular, simulations reported below are realized using $B = 0.35$ T, $D = -0.81$ cm$^{-1}$ and $E = -0.24$ cm$^{-1}$ for $S = 3/2$. Then, by varying *S*, the parameters in Hamiltonian (8*b*) are re-scaled by the common factor $1/S(2S-1)$ to keep the energy gaps involved in the protocol (and hence the frequency of the pulses used for manipulations) within experimental capabilities $\lesssim 24$ GHz. This re-scaling is the same observed for the zero-field splitting terms in the ground state of multi-spin single-molecule magnets [53].

These parameters are typical of existing compounds, such as single spin *S* complexes like the spin 3/2 Cr-based qudit [Cr(C$_3$S$_5$)$_3$]$_3$ [50] or the spin 7/2 GdW$_{30}$ [51]. Both these complexes display a high rhombicity $E/D$ between 0.2 and 0.3, a feature which ensures connectivity between all pairs of system eigenstates and hence efficient implementation of the protocol. Nonetheless, this choice of parameters is not critical: smaller $E/D$ values would slightly slow down the control of the system, without compromising the fault-tolerance of the scheme [18]. Moreover, the symmetry of the system can be decreased from axial to rhombic by distortion of the ligand field. Finally, we point out that a wide range of molecules could fit our scheme. Besides complexes containing a single magnetic ion, multi-spin clusters consisting of several interacting spins with a spin *S* ground multiplet can be used. This class of molecules, also known as single-molecule magnets, were extensively studied in the last decades for their slow relaxation dynamics suitable for single molecule information storage. To this aim, many molecules were synthesized [54] with total spin *S* reaching large values (up to 10–20) and also suitable to deposition on surfaces, a property which would ease an experimental implementation [55, 56].

Concerning the actual realization of a device, we note that molecular spin qudits can be controlled by embedding them into superconducting resonators, and then adapting strategies developed for the control of superconducting qubits. A detailed description of this setup, with a blueprint of the molecular spin quantum processor, can be found in [57]. Strong coupling between individual molecules and photons in the resonator can be accomplished by designing proper constrictions in the resonator in which the magnetic field is concentrated and hence the coupling is magnified. Together with the use of molecules with large total spin *S*, this results in an enhancement of the spin-photon interaction of several orders of magnitude. The strong coupling can then be exploited to read-out the spin qudit state, while classical microwave drives sent through control lines are exploited to manipulate the qudit state.

### 3.3. Protected code words

Since $H_d^{(1)}$ is only a very weak perturbation (with unknown $B_x$), we derive protected code words starting from the eigenstates of $H_d^{(0)}$ and considering dephasing errors acting on them.

Pure dephasing can be modeled by error operators *diagonal* in the system eigenstates $|\mu\rangle$, $\mu = 0, \ldots, 2S$ (labeled in order of increasing energy). For spin *S* MNMs, the coupling with the environment occurs via magnetic dipole operators $S_\alpha$. Hence, the *diagonal* $E_k$ describing pure dephasing are in general represented by powers of $S_\alpha$ [35, 58]. Here, the form of Hamiltonian (8*a*) (typical of many single-ion and single molecule magnets [54]) yields precise selection rules, with spin eigenstates which are only superpositions of $\Delta m = \pm 2$ states ($S_z|m\rangle = m|m\rangle$). As a result, only $S_z$ has diagonal matrix elements in the eigenbasis, $\langle\mu|S_\alpha|\mu\rangle \propto \delta_{\alpha,z}$. Moreover, $S_{x,y}$ and $S_z$ operators connect different eigenstates, as schematically shown in figure 3(a) for a spin $S = 3/2$.

A convenient choice for defining a QEC code is based on selecting disjoint subspaces for logical states $\ell = 0, 1$. In the illustrative $S = 3/2$ case shown in figure 3, this corresponds to encode $|\ell = 0, k\rangle$ into a superposition of eigenstates $|\mu = 0\rangle$ and $|\mu = 3\rangle$ (red bars in figure 3(b)) and $|1, k\rangle$ into a superposition of $|\mu = 1\rangle$ and $|\mu = 2\rangle$ (blue bars). The coefficients of this superpositions are determined numerically to satisfy Knill–Laflamme conditions (see appendix F).

This choice of the code/error words ensures that $S_x$ has matrix elements $\langle 0, k|S_x|1, k\rangle$ (see figure 3(a)) between different logical subspaces $\ell = 0, 1$ with the same index *k*, thus potentially activating logical Rabi oscillations if the proper energy gaps are matched by longitudinal drives. Moreover (figure 3(a)), $S_z$ has both diagonal matrix elements (exploited to activate the Rabi, see section 3.4 and appendix B) and off-diagonal ones between eigenstates defining the same code/error word ($\langle\ell, k|S_z|\ell, k'\rangle \neq 0$), a condition which allows for stabilization and correction (see section 3.5 and appendix D). As further stressed below, the logical rotation is activated by a term proportional to the *transverse* field to be measured, while all manipulations needed to apply error correction are done via matrix elements of the *longitudinal* oscillating field. The corresponding matrix elements of $S_x$ and $S_z$ do not overlap (see figure 3(a) for $S = 3/2$), and this ensures that sensing and error correction procedure are completely independent, thus making the signal clearly distinguishable from the noise.

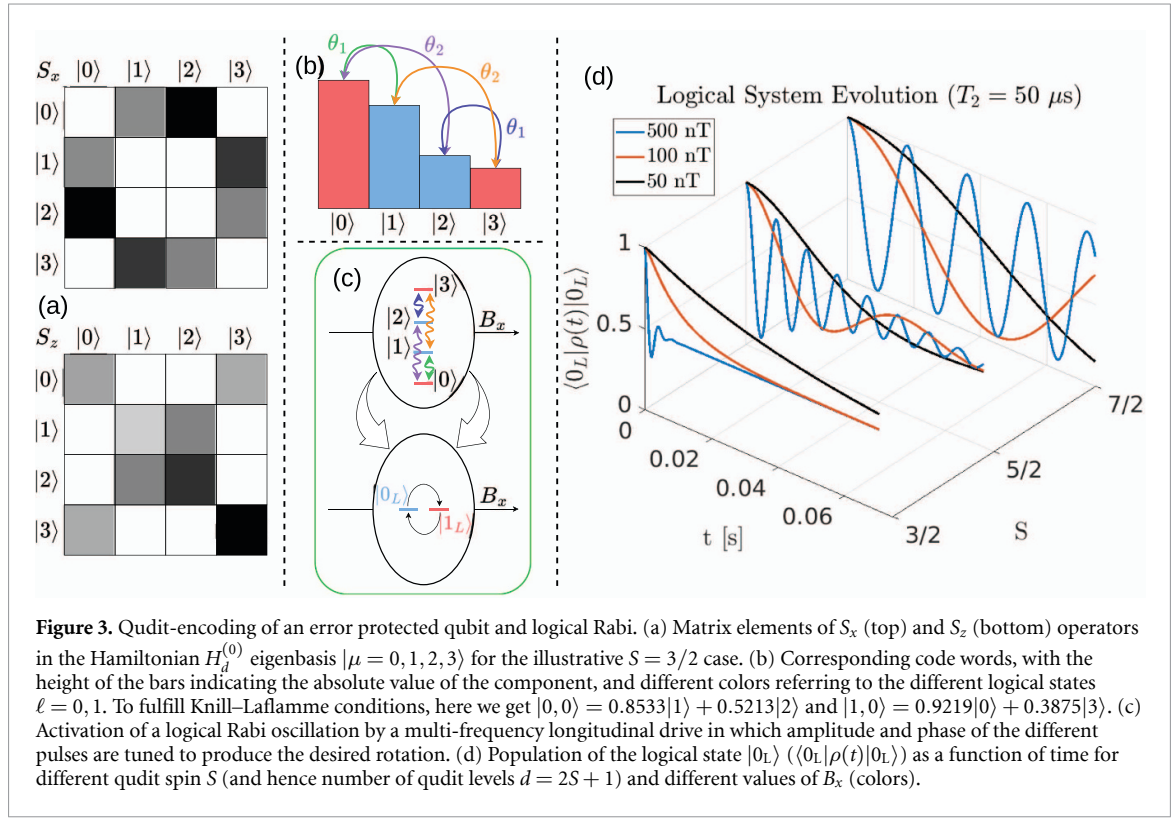

**Figure 3.** Qudit-encoding of an error protected qubit and logical Rabi. (a) Matrix elements of $S_x$ (top) and $S_z$ (bottom) operators in the Hamiltonian $H_d^{(0)}$ eigenbasis $|\mu = 0,1,2,3\rangle$ for the illustrative $S = 3/2$ case. (b) Corresponding code words, with the height of the bars indicating the absolute value of the component, and different colors referring to the different logical states $\ell = 0,1$. To fulfill Knill–Laflamme conditions, here we get $|0,0\rangle = 0.8533|1\rangle + 0.5213|2\rangle$ and $|1,0\rangle = 0.9219|0\rangle + 0.3875|3\rangle$. (c) Activation of a logical Rabi oscillation by a multi-frequency longitudinal drive in which amplitude and phase of the different pulses are tuned to produce the desired rotation. (d) Population of the logical state $|0_L\rangle$ ($\langle 0_L|\rho(t)|0_L\rangle$) as a function of time for different qudit spin $S$ (and hence number of qudit levels $d = 2S+1$) and different values of $B_x$ (colors).

### 3.4. Logical Rabi oscillation

A logical Rabi oscillation is defined as follows

$$R_x^L(\vartheta) = \cos\frac{\vartheta}{2}\sum_k (|0,k\rangle\langle 0,k| + |1,k\rangle\langle 1,k|)$$
$$- i\sin\frac{\vartheta}{2}\sum_k (|0,k\rangle\langle 1,k| + |1,k\rangle\langle 0,k|). \quad (9)$$

In order to activate $R_x^L(\vartheta)$ by $H_d^{(1)}$, we need to simultaneously address all transitions between $|\ell = 0,k\rangle$ and $|\ell = 1,k\rangle$ error words, as sketched in figure 3(a). We do this by a longitudinal multi-frequency oscillating field of the form

$$h_d^1(t) = g\mu_B S_z \sum_j b_{z,j}\cos(\omega_{z,j}t + \phi_j). \quad (10)$$

The effect of $h_d^1(t)$ can be understood by generalizing the transformation to the rotating frame applied to the qubit case, as detailed in appendix B. In particular, we introduce the transformation $U(t) = \exp[ig\mu_B\sum_j b_{z,j}/\omega_{z,j}\sin(\omega_{z,j} + \phi_j)S_z^{(D)}]$, thus obtaining the *qudit* rotating-frame Hamiltonian $H_{rf}^d = U(t)[H_d^{(0)} + H_d^{(1)} + h_d^1(t)]U^\dagger(t) + i\,\dot{U}(t)U^\dagger(t)$. In the perturbative $g\mu_B b_{z,j}/\omega_{z,j} \ll 1$ limit, this reduces to

$$H_{rf}^d \approx H_0 + i\,(g\mu_b)^2\sum_j \frac{B_x b_{z,j}}{\omega_{z,j}}\sin(\omega_{z,j}t + \phi_j)\left[S_z^{(D)}, S_x\right] + g\mu_B\sum_j b_{z,j}S_z^{(OD)}\cos(\omega_{z,j}t + \phi_j). \quad (11)$$

Here we have separated the diagonal $[S_z^{(D)}]$ and off-diagonal $[S_z^{(OD)}]$ parts of $S_z$ (see sketch in figure 3(a)), due to their different effect on the system dynamics.

The transformed Hamiltonian includes a multi-frequency transverse driving field with re-normalized amplitudes $B_x b_{z,j}/\omega_{z,j}$, besides an additional drive $\propto b_{z,j}S_z^{(OD)}$. Remarkably, the two oscillating terms connect different eigenstates, exactly as the $S_x$ and $S_z$ matrices of figure 3(a), and are exploited for the logical Rabi and for the error correction, respectively.

In the limit of small $g\mu_B b_{z,j}/\omega_{z,j}$ we are considering here, each transition between a pair of eigenstates $|\mu\rangle$ and $|\nu\rangle$ of $H_d^{(0)}$ is activated by tuning the frequency of the drive in resonance with the corresponding energy gap, i.e. setting $\omega_{z,j} = E_\mu - E_\nu$. All other off-diagonal terms in the rotating frame Hamiltonian (appendix B) are ineffective, since they are largely off-resonance. The Rabi frequency of the oscillation induced by the first term in equation (11) is proportional to $B_x|\langle\mu|[S_z^{(D)}, S_x]|\nu\rangle|b_{z,j}/|E_\mu - E_\nu|$ (see appendix B). Hence, by simultaneously sending several resonant pulses we activate Rabi oscillations between different pairs of eigenstates and, in particular, between each component of $\ell = 0$ and each component of $\ell = 1$ (figures 3(b) and (c)). The period of these oscillations can be tuned by choosing the value of $b_{z,j}$ in order to result in a *logical Rabi* between the code words $|0_L\rangle$ and $|1_L\rangle$.

The use of a multi-frequency drive $h_d^1(t)$ requires attention when addressing transitions involving not-independent pairs of states, such as $|\mu=0\rangle \to |\mu=1\rangle$ and $|\mu=1\rangle \to |\mu=2\rangle$ in the 4-level case shown in figure 3. Indeed, in this case an unwanted transition between $|\mu=0\rangle \to |\mu=2\rangle$ could be triggered, as emerges from the form of the generalized rotating frame Hamiltonian and discussed in detail in appendix B. This hurdle can be overcome by resorting to a Suzuki–Trotter decomposition of the time evolution (see appendix C). This implies separating the multi-frequency drive into independent subsets of short pulses which are sent in sequence. The resulting error in the implementation of $R_x^L(\vartheta)$ can be made arbitrary small by a sufficiently large number of steps, without affecting the duration of the logical gate nor the efficiency of the sensing protocol (see simulations in appendix C).

### 3.5. Error detection and correction

Following the scheme developed in [18], our method relies on using a $d/2$-level ancilla in conjunction with the $d$-level LQ for detecting errors. Suppose the exact state of the computation is $\alpha|0,0\rangle + \beta|0,0\rangle$ and hence the corresponding encoded state after possible errors is $|\bar\psi\rangle = \sum_k [\alpha_k|0,k\rangle + \beta_k|1,k\rangle])$. Then the error detection step consists in the execution of a *k*-controlled operation, denoted as $\mathscr{S}$, between the LQ $|\bar\psi\rangle$ and the ancilla (serving as target and initialized in its ground state $|0\rangle$). The $\mathscr{S}$ gate operates as follows: it maps $|\ell,k\rangle|0\rangle$ to $|\ell,k\rangle|k\rangle$; $|\ell,k\rangle|k\rangle$ to $-|\ell,k\rangle|0\rangle$, and it is the identity on any other states [18]. Consequently, by applying the gate $\mathscr{S}$ to the system in the state $|\bar\psi\rangle|0\rangle$ we map it to $\sum_k[\alpha_k|0,k\rangle + \beta_k|0,k\rangle]|k\rangle$. Hence, $\mathscr{S}$ correlates each error $k$ with a specific eigenstate of the ancilla. Subsequently, by measuring the ancilla in its eigenbasis, the error syndrome $\bar k$ can be identified, thereby stabilizing information within the $|\ell,\bar k\rangle$ subspace. Indeed, if the measurement of the ancilla has given $\bar k$ as a result, the state of the qudit is $\tilde\alpha_{\bar k}|0,\bar k\rangle + \tilde\beta_{\bar k}|1,\bar k\rangle$, where $\tilde\alpha_{\bar k}$ and $\tilde\beta_{\bar k}$ are the renormalized coefficient of the superposition. Fulfilling the Knill–Laflamme conditions ensures that such coefficients are close to the ones of the ideal superposition, $\alpha$ and $\beta$.

Finally, the correction step involves reverting the stabilized states $\bar\alpha_{\bar k}|0,\bar k\rangle + \bar\beta_{\bar k}|1,\bar k\rangle$ back to $\bar\alpha_{\bar k}|0,0\rangle + \bar\beta_{\bar k}|1,0\rangle$. This is done by the operator $\zeta_{\bar k}$ that maps $|\ell,\bar k\rangle$ to $|\ell,0\rangle$ and vice versa [18].

Both the $\mathscr{S}$ gate and the recovery operation $\zeta_k$ require resonant pulses addressing transitions between eigenstates *within the same code/error word* [18]. This is achieved by a driving field along $z$. Indeed, besides the diagonal matrix elements exploited to activate the Rabi oscillation, $S_z$ displays off-diagonal matrix elements between $|\ell,k\rangle$ and $|\ell,k'\rangle$, with $\langle\ell,k|S_z|\ell',k'\rangle \propto \delta_{\ell,\ell'}$ (see figure 3(a) for $S=3/2$). Note that diagonal elements on the driving field here only yield additional phases, which can be properly compensated.

This peculiar feature of Hamiltonian (8*a*), yielding $S_z$ and $S_x$ with off-diagonal matrix elements between different pairs of eigenstates, allows us to separate Logical Rabi oscillations from error correction steps in the sensing protocol, thus significantly simplifying it. Moreover, this ensures that logical rotations and the error correction procedure are independent and hence the signal is completely distinguishable from the noise to be corrected. We remark again that this feature can be easily obtained in MNMs, since Hamiltonian (8*a*) is typical of many single-ion and single-molecule magnets [54] largely studied in the field.

A final comment concerns the measurement of the state of the system. As already explained above, for the error detection step we need to measure the ancilla in its eigenbasis. Measurement of a single molecular spin can be achieved by coupling it to a superconducting resonator, as discussed in section 3.2 and in the detailed blueprint presented in [57]. Conversely, determining the state of the LQ after a partial Rabi oscillation involves distinguishing the value of $\ell$ regardless of the error $k$. Since $\ell = 0$ and $\ell = 1$ are encoded into disjoint subspaces, the readout simply entails summing the probabilities of finding the LQ in the eigenstates belonging to each of these two subspaces.

## 4. Results

Logical Rabi oscillations obtained by numerical solution of the Lindblad equation (5) for a spin *S*, including dephasing and evolution induced by the Hamiltonian in the rotating frame (equation (B11)), are reported in

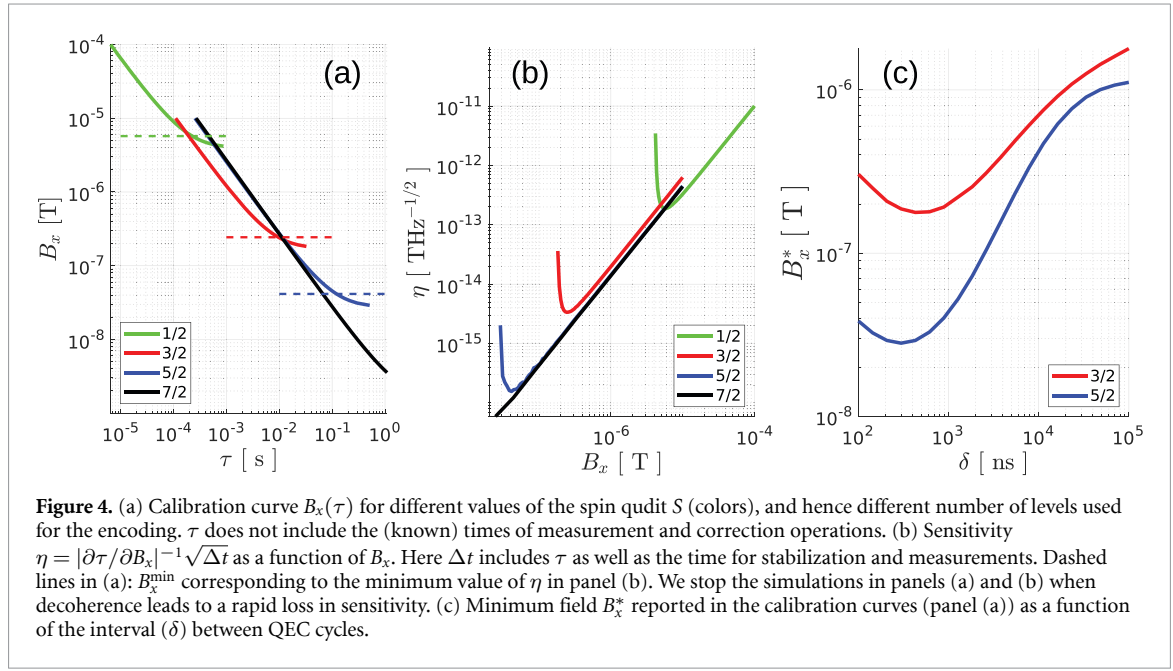

**Figure 4.** (a) Calibration curve $B_x(\tau)$ for different values of the spin qudit $S$ (colors), and hence different number of levels used for the encoding. $\tau$ does not include the (known) times of measurement and correction operations. (b) Sensitivity $\eta = |\partial\tau/\partial B_x|^{-1}\sqrt{\Delta t}$ as a function of $B_x$. Here $\Delta t$ includes $\tau$ as well as the time for stabilization and measurements. Dashed lines in (a): $B_x^{\min}$ corresponding to the minimum value of $\eta$ in panel (b). We stop the simulations in panels (a) and (b) when decoherence leads to a rapid loss in sensitivity. (c) Minimum field $B_x^*$ reported in the calibration curves (panel (a)) as a function of the interval ($\delta$) between QEC cycles.

figure 3(d). Error correction has been applied each $\delta = 500$ ns and leakage errors, which could arise from the finite spectral width of the pulses, have been neglected. The latter is a reasonable assumption, since these errors are strongly system-specific and can be largely reduced by pulse-shaping techniques [59] which we are not considering here.

Compared with the qubit case without error correction, we immediately note a huge enhancement of the effective $T_2$, which increases by increasing $S$. For $S = 7/2$ the decay of Rabi oscillations occurs on a time-scale $\sim 0.5$ s $= 10^4 \, T_2$, i.e. the effective $T_2$ is amplified by roughly 4 orders of magnitude. As a result, much smaller transverse fields can be detected. In particular, $B_x$ as small as 50 nT still leads to clearly visible oscillations for a $S = 7/2$ qudit.

The remarkable improvement brought by QEC to the sensing protocol can be quantified by considering the calibration curve $B_x(\tau)$ and the corresponding sensitivity $\eta$, in analogy to section 2. Results are reported in figure 4 for $T_2 = 50\,\mu$s, reasonable for molecular spins [41, 52]. The performance of our protocol improves with the qudit spin $S$: both $B_x^{\min}$ and the corresponding minimum sensitivity (figures 4(a) and (b)) decrease by increasing $S$: from the $\sim 6\,\mu$T of the qubit to $\sim 0.2\,\mu$T for $S = 3/2$, $\sim 40$ nT for $S = 5/2$ and a few nT for $S = 7/2$ ($B_x^{\min}$ is below numerical accuracy). Already with a spin 3/2 qudit, these values are of the order of the minimum detectable field for the Ramsey protocol $\sim \frac{\hbar}{g\mu_b T_2} \sim 0.1\,\mu$T [47]. Thanks to the proposed FT approach, the use of spin qudits with larger $S$ enables the detection of much smaller transverse fields, no longer limited by $T_2$.

Finally, we consider the effectiveness of our scheme as a function of the frequency of the time interval ($\delta$) between QEC cycles (figure 4(c)). The best performance is achieved by using $\delta^* \approx 400$ ns. Nonetheless, the performance of the scheme remains remarkable even for larger $\delta \sim 2\,\mu$s. The performance decreases for $\delta > \delta^*$ due to an accumulation of errors not completely corrected by the code, which increases at a polynomial rate with $\delta$ [18]. At the same time, reducing $\delta$ implies increasing the number of QEC cycles (each of a fixed duration) and hence of the total processing time. This, in turn, yields an accumulation of errors not completely corrected by the code, resulting in a worsening of the protocol performance when too many QEC cycles are applied. Nevertheless, $\delta^*$ shortens as $S$ increases, due to the increasing correcting power of the code. As a consequence, the uncorrected errors that accumulate during a QEC cycle decrease by increasing $S$.

## 5. Discussion and conclusions

We propose a FT scheme for quantum sensing of transverse magnetic fields based on *individual molecular spin qudits* acting as quantum sensors. By embedding QEC within each unit and implementing it fault-tolerantly, we numerically demonstrate the capability to detect tiny magnetic fields, not limited by $T_2$ and without introducing any systematic bias.

The scheme is based on designing the system Hamiltonian and the pulse sequence in such a way that the signal generates a *FT logical Rabi rotation* between encoded states, whose frequency is *linear* in the measured quantity. This is achieved by exploiting the possibility to drive transitions between each pair of eigenstates of the qudit sensor, activated either by the transverse field to be probed (in combination with a longitudinal oscillating field) or by proper resonant pulses.

The above requirements are easily met by a wide class of MNMs. These clusters stand out for the high-degree of control at the synthetic level, which translates into a high tunability of their spin Hamiltonian. In particular, simple MNMs such as spin *S* molecules in a low-symmetry zero-field splitting are ideal for our scheme [23, 24, 54]. Remarkably, several molecules of this kind already exist [50, 51] and could be used for the first experimental implementations of the protocol (see section 3.2). Control and readout of the system could be accomplished by placing molecules within superconducting resonators, as detailed in the realistic blueprint of a molecular spin quantum processor presented in [57].

Other experimental platforms with remarkable coherence could be envisaged to implement our protocol, such as ultracold atoms [4, 60–62] and trapped ions [63–66]. There, (hyper-)fine multiplets of alkaline, earth-alkaline or lanthanides atoms could be exploited, if the proper inter-level connectivity is provided.

Compared to other schemes for QEC exploiting many qubits, the present *qudit approach* encodes a protected qubit within a single object [28, 32]. Remarkably, the performance we obtain in the enhancement of $T_2$ with a small number of qudit levels would require hundreds of qubits with standard methods [18]. As a result, the qudit approach greatly simplifies manipulations, because many error-prone multi-qubit gates are replaced by two-level rotations within the same object, much easier to implement. Moreover, replacing a large collection of qubits with a single qudit enables a much higher spatial resolution.

Some points deserve a discussion. The first one concerns the errors handled by our scheme. It is important to note that, although we focused on pure dephasing (which is by far the leading one at low temperatures in molecular spin clusters), the FT scheme for quantum sensing presented here can be adapted to other errors. For instance, relaxation errors could be faced by increasing the number of qudit levels and then adapting the code, without altering the FT scheme for quantum sensing. Indeed, relaxation could become relevant for the long measurement times ($10^{-2}$–$10^{-1}$ s) required to detect very small values of $B_x$. Nonetheless, spin relaxation times of up to 1 s were reported for molecular spins at ∼7 K [49], which allow for the application of the protocol in its present form without compromising significantly its performance (see appendix E). Moreover, the relaxation time should further increase by decreasing temperature to the sub-K range typical of quantum information processing.

Finally, finite rate QEC can introduce an estimation bias when applied to quantum sensing, thus reducing the sensor's performance. It was shown [15] that this systematic error can be prevented under specific conditions. In particular, our FT procedure does not induce any bias, because it yields the same time evolution for both logic- and error-subspaces under the applied unitaries [15].

## Data availability statement

The data that support the findings of this study are openly available at the following URL/DOI: https://doi.org/10.5281/zenodo.12634522.

## Acknowledgments

We warmly thank Paolo Santini, Augusto Smerzi, and Sandro Wimberger for useful and stimulating discussions.

The work received funding from the European Union—NextGenerationEU, PNRR MUR Project PE0000023-NQSTI, from Novo Nordisk foundation under Grant NNF21OC0070832 in the call 'Exploratory Interdisciplinary Synergy Programme 2021' and from Fondazione Cariparma. MM acknowledges funding from the European Union—NextGenerationEU under the National Recovery and Resilience Plan (NRRP), Mission 4 Component 1 Investment 3.4 and 4.1. Decree by the Italian Ministry N. 351/2022 CUP D92B22000530005.

## Appendix A. Derivation of the qubit Hamiltonian in equation (3)

In this appendix, we derive equation (3) of the main text, as well as some interesting generalizations of it. We start from the Hamiltonian

$$H(t) = \hbar\Omega\,\sigma_z + g\mu_B\left(B_x + B_{1x}\cos\left(\omega_x t + \phi_x\right)\right)\sigma_x + g\mu_B B_{1z}\cos\left(\omega_z t + \phi_z\right)\sigma_z \quad (A1)$$

($\mu_B$ being the Bohr magneton and $g \approx 2$ the *g*-factor), describing a two-level system ($s = \frac{1}{2}$), diagonal in the energy eigenstates basis (we neglected a possible contribute proportional to the identity), minimally coupled to a constant magnetic field $B_x$ along $\hat{x}$, and to two oscillating magnetic fields $B_{1x}$ and $B_{1z}$. We perform the unitary transformation

$$U(t) = e^{i \frac{g\mu_b B_{1z}}{\hbar \omega_z} \sin(\omega_z t + \phi_z) \sigma_z}, \tag{A2}$$

diagonal in the basis of the energy eigenstates. In this way,

$$H'(t) = i\hbar \, \partial_t U(t) \, U(t)^{-1} + U(t) H(t) U(t)^{-1}, \tag{A3}$$

so that explicitly we obtain:

$$H'(t) = \hbar \Omega \, \sigma_z + [g\mu_b (B_x + B_{1x} \cos(\omega_x t + \phi_x)) \sigma_x] \, e^{-i \left[ 2 \frac{g\mu_b B_{1z}}{\hbar \omega_z} \sin(\omega_z t + \phi_z) \right] \sigma_z}, \tag{A4}$$

where the part $g\mu_b B_{1z} \cos(\omega_z t + \phi_z) \sigma_z$ has been canceled out by the part $i \partial_t U(t) U(t)^{-1}$. The latter expression can be managed, noticing that

$$e^{-i2\left(\frac{g\mu_b B_{1z}}{\hbar \omega_z}\right) \sin(\omega_z t + \phi_z) \sigma_z} = \cos\left[ 2 \left( \frac{g\mu_b B_{1z}}{\hbar \omega_z} \right) \sin(\omega_z t + \phi_z) \right] \mathbf{I} - i \sin\left[ 2 \left( \frac{g\mu_b B_{1z}}{\hbar \omega_z} \right) \sin(\omega_z t + \phi_z) \right] \sigma_z, \tag{A5}$$

Moreover, in the limit $g\mu_B B_{1z} \ll \hbar \omega_z$ we obtain

$$\cos\left[ \left( 2 \frac{g\mu_b B_{1z}}{\hbar \omega_z} \right) \sin(\omega_z t + \phi_z) \right] = \frac{1}{2} \left[ \left( 1 - \cos\left( 2 \frac{g\mu_b B_{1z}}{\hbar \omega_z} \right) \right) \cos(\omega_z t + \phi_z) + \left( 1 + \cos\left( 2 \frac{g\mu_b B_{1z}}{\hbar \omega_z} \right) \right) \right], \tag{A6}$$

and

$$\sin\left[ 2 \left( \frac{g\mu_b B_{1z}}{\hbar \omega_z} \right) \sin(\omega_z t + \phi_z) \right] = \sin\left( 2 \frac{g\mu_b B_{1z}}{\hbar \omega_z} \right) \sin(\omega_z t + \phi_z). \tag{A7}$$

Actually, they are a good approximation provided $g\mu_B B_{1z} < \hbar \omega_z$. Exploiting these relations and the standard trigonometric prosthaphaeresis identities, we finally obtain:

$$\begin{aligned} H'(t) \approx &\ \hbar \Omega \sigma_z + \frac{g\mu_b B_x}{2} \left( 1 + \cos\left( 2 \frac{g\mu_b B_{1z}}{\hbar \omega_z} \right) \right) \sigma_x + \frac{g\mu_b B_x}{2} \left( 1 - \cos\left( 2 \frac{g\mu_b B_{1z}}{\hbar \omega_z} \right) \right) \cos 2(\omega_z t + \phi_z) \sigma_x \\ &- g\mu_b B_x \sin\left( 2 \frac{g\mu_b B_{1z}}{\hbar \omega_z} \right) \sin(\omega_z t + \phi_z) \sigma_y + \frac{g\mu_b B_{1x}}{2} \left( 1 + \cos\left( 2 \frac{g\mu_b B_{1z}}{\hbar \omega_z} \right) \right) \cos(\omega_x t + \phi_x) \sigma_x \\ &+ \frac{g\mu_b B_{1x}}{4} \left( 1 - \cos\left( 2 \frac{g\mu_b B_{1z}}{\hbar \omega_z} \right) \right) \left[ \cos((2\omega_z + \omega_x)t + (2\phi_z + \phi_x)) + \cos((2\omega_z - \omega_x)t + (2\phi_z - \phi_x)) \right] \sigma_x \\ &- \frac{g\mu_b B_{1x}}{2} \sin\left( 2 \frac{g\mu_b B_{1z}}{\hbar \omega_z} \right) \left[ \sin((\omega_z + \omega_x)t + (\phi_z + \phi_x)) + \sin((\omega_z - \omega_x)t + (\phi_z - \phi_x)) \right] \sigma_y, \end{aligned} \tag{A8}$$

or, expanding in the small parameter $\frac{g\mu_b B_{1z}}{\hbar \omega_z} \ll 1$:

$$\begin{aligned} H'(t) \approx &\ \hbar \Omega \sigma_z + g\mu_b B_x \sigma_x - \frac{2 (g\mu_b)^2}{\hbar} \frac{B_x B_{1z}}{\omega_z} \sin(\omega_z t + \phi_z) \sigma_y + g\mu_b B_{1x} \cos(\omega_x t + \phi_x) \sigma_x \\ &- \left( \frac{(g\mu_b)^2}{\hbar} \frac{B_{1x} B_{1z}}{\omega_z} \right) \left[ \sin((\omega_z + \omega_x)t + (\phi_z + \phi_x)) + \sin((\omega_z - \omega_x)t + (\phi_z - \phi_x)) \right] \sigma_y. \end{aligned} \tag{A9}$$

If further $\omega_x = \omega_z = \omega$, and $\phi_x = \phi_z = \phi$ (that means from a unique pulse, along $\hat{x}$ and $\hat{z}$), the last expression reduces to:

$$\begin{aligned} H'(t) \approx &\ \hbar \Omega \sigma_z + g\mu_b B_x \sigma_x - 2 \frac{(g\mu_b)^2}{\hbar} \frac{B_x B_{1z}}{\omega} \sin(\omega t + \phi) \sigma_y + g\mu_b B_{1x} \cos(\omega t + \phi) \sigma_x \\ &- \frac{(g\mu_b)^2}{\hbar} \frac{B_{1x} B_{1z}}{\omega} \sin 2(\omega t + \phi) \sigma_y. \end{aligned} \tag{A10}$$

Instead, if $B_{1x} = 0$, the latter two equations reduce to

$$H'(t) \approx \hbar \Omega \sigma_z + g\mu_b B_x \sigma_x - \frac{(g\mu_b)^2}{\hbar} \frac{2 B_x B_{1z}}{\omega_z} \sin(\omega_z t + \phi_z) \sigma_y. \tag{A11}$$

In the latter expression, the second term renormalizes the energy levels $\pm\hbar\Omega$ by an amount $\sim(g\mu_b B_x)^2$ but its contribution can be neglected, since we assume the energy $g\mu_b B_x$ much lower than any other energy scale in the problem (this is true for the values of $B_x$ that we propose to estimate). Neglecting this term, discarding the counter-propagating wave $e^{-i\omega_z t}$, and moving to a rotating frame by the transformation $|\downarrow\rangle \to i|\downarrow\rangle e^{i\frac{E_\downarrow}{\hbar}t}$, the previous Hamiltonian can be rephrased to the time-independent form:

$$H' \approx -\frac{(g\mu_b)^2}{\hbar}\frac{B_x B_{1z}}{\omega_z}\sigma_x, \tag{A12}$$

equivalent to equation (3) in the main text at resonance, provided to set $\hbar \equiv 1$, $\Omega = \Delta/2$ and to recall that $s_\alpha = \sigma_\alpha/2$.

## Appendix B. Derivation of the Hamiltonian for a qudit in equation (11).

In this appendix, we discuss the generalization of some results of the previous appendix, concerning a two-level system, to $S > 1$. We start from Hamiltonian (8*b*), that is

$$H(t) = H_0 + g\mu_b B_x S_x + g\mu_b \sum_j b_{z,j} \cos\left(\omega_{z,j} t + \phi_{z,j}\right) S_z. \tag{B1}$$

In the following, it will be useful to explicitly separate $S_z$ in diagonal and not-diagonal parts: $S_z = S_z^{(D)} + S_z^{(OD)}$. Consequently and similarly to equation (A2), the strategy that we perform is to eliminate the time-dependent term of equation (B1) proportional to $S_z^{(D)}$, via the time-dependent unitary transformation

$$U(t) = e^{ig\mu_b \sum_j \frac{b_{z,j}}{\hbar\omega_{z,j}} \sin\left(\omega_{z,j} t + \phi_{z,j}\right) S_z^{(D)}}. \tag{B2}$$

Under this transformation, $H(t)$ transforms as

$$H'(t) = i\hbar\partial_t U(t)\, U^{-1}(t) + U(t) H(t) U^{-1}(t), \tag{B3}$$

and explicitly:

$$H'(t) = H_0 + U(t)\left(g\mu_b B_x S_x + g\mu_b \sum_j b_{z,j} \cos\left(\omega_{z,j} t + \phi_{z,j}\right) S_z^{(OD)}\right) U^{-1}(t). \tag{B4}$$

This expression is still exact. Notice that in the latter Hamiltonian the term $\propto S_z^{(OD)}$, originally present in equation (B1), does not appear any longer, canceled out by the first term in equation (B3).

The latter Hamiltonian cannot be simplified further exactly, as for the spin-$\frac{1}{2}$ case, since a counterpart of equation (A6) is not available. A solution based on the expansion of the exponential in $U(t)$ in terms of Bessel function has been provided in [19]. The same form is not particularly manageable for our purposes. Instead, we adopt again a perturbative approach since $\eta_j \equiv \frac{g\mu_b b_{z,j}}{\hbar\omega_{z,j}} \ll 1$. In this perturbative limit, the exponential factors in equation (B4) can be expanded, obtaining, at the leading order in the perturbative parameters $\eta_j$:

$$e^{ig\mu_b \sum_j \frac{b_{z,j}}{\hbar\omega_{z,j}} \sin(\omega_{z,j} t + \phi_{z,j}) S_z^{(D)}} = \mathbf{I} + ig\mu_b \sum_j \frac{b_{z,j}}{\hbar\omega_{z,j}} \sin\left(\omega_{z,j} t + \phi_{z,j}\right) S_z^{(D)} + o\left(\{\eta_j\}\right). \tag{B5}$$

In this way, again at the leading order in $\eta_j$:

$$\begin{aligned}H'(t) \approx & \left(H_0 + g\mu_b B_x S_x + g\mu_b \sum_j b_{z,j} \cos\left(\omega_{z,j} t + \phi_{z,j}\right) S_z^{(OD)}\right) \\ & + ig\mu_b \sum_j \frac{b_{z,j}}{\hbar\omega_{z,j}} \sin\left(\omega_{z,j} t + \phi_{z,j}\right) \left(g\mu_b B_x \left[S_z^{(D)}, S_x\right] + g\mu_b \sum_l b_{z,l} \cos\left(\omega_{z,l} t + \phi_{z,l}\right) \left[S_z^{(D)}, S_z^{(OD)}\right]\right).\end{aligned} \tag{B6}$$

The second term shifts the eigenvalues of $H_0$ by contributes $\propto B_x^2$ and can therefore be neglected. The fourth term in equation (B6), proportional to $B_x$, implements oscillations between logical states, allowing to estimate $B_x$ itself. As discussed in the main text and in [18], we can easily choose the Hamiltonian parameters of equation (B6) such that the matrix elements of this term connect logical and error- states separately. The Rabi periods from this term are (still in the absence of decoherence)

$$\tau_j = \frac{\hbar}{\Omega_{z,j}} = \frac{\hbar}{(g\mu_b)^2} \frac{|E_\mu - E_\nu|}{b_{z,j} B_x} \frac{1}{\langle \mu | \left[ S_z^{(D)}, S_x \right] | \nu \rangle}, \tag{B7}$$

where we have set the resonance condition $E_\mu - E_\nu = \hbar\omega_{z,j}$. Notably, the (shifted) periods must be set all equal each others, $\equiv \tau$, along the logic evolution, choosing properly the set $\{b_{z,j}\}$.

The effect of the last term in equation (B6) can be understood by exploiting the trigonometric identities:

$$\sin(\omega_{z,j}t + \phi_{z,j})\cos(\omega_{z,l}t + \phi_{z,l}) = \frac{1}{2}\left[\sin\left((\omega_{z,j} + \omega_{z,l})t + \phi_{z,j} + \phi_{z,l}\right) + \sin\left((\omega_{z,j} - \omega_{z,l})t + \phi_{z,j} - \phi_{z,l}\right)\right]. \tag{B8}$$

Clearly, certain combinations of frequencies $(\omega_{z,j} \pm \omega_{z,l})$ amount to frequencies relative to resonant (multi-photon) transitions that we do not want to induce. Indeed, $[S_z^{(D)}, S_z^{(OD)}]$ has exactly the matrix elements relative to these transitions, as $S_z^{(OD)}$ itself in the third term in equation (B6) (however the transitions from this term are excluded, since they are not resonant).

A strategy to avoid this problem in a realistic implementation is to adopt a Trotter decomposition approach. Specifically, this involves organizing the couplings $g\mu_b b_{z,j} \cos(\omega_{z,j}t + \phi_{z,j})S_z$ to act on $H_0 + g\mu_b B_x S_x$. The various subsets act separately from each other over time.

It is immediate to check that, if the division in pairs is performed suitably, the unwanted resonant transitions from some of the combinations $(\omega_{z,j} \pm \omega_{z,l})$, $j \neq l$, are avoided. In this way, all the contributes from the terms

$$i(g\mu_b)^2 \sum_{j,l} \frac{b_{z,j} b_{z,l}}{\hbar\omega_{z,j}} \sin(\omega_{z,j}t + \phi_{z,j})\cos(\omega_{z,l}t + \phi_{z,l})\left[S_z^{(D)}, S_z^{(OD)}\right] \tag{B9}$$

effectively result in shifts of the resonances that we exploit for the logic operations.

Similar shifts occur from the third term in equation (B6); indeed, it can be checked that the nonvanishing matrix elements from the commutators are in the same positions.

It is important to note that since $g\mu_b b_{z,j} \ll 1$ and $\frac{(g\mu_b)^2 b_{z,j} b_{z,l}}{\hbar\omega_{z,j}} \ll 1$ all the aforementioned shifts are negligible. Thus we can write the effective Hamiltonian as

$$H_I'^{(\text{red})}(t) \approx i(g\mu_b)^2 \sum_{\alpha,j\in\alpha} \frac{B_x b_{z,j}}{\hbar\omega_{z,j}} \sin(\omega_{z,j}t + \phi_{z,j})\left[S_z^{(D)}, S_x\right], \tag{B10}$$

retrieving the formulation reported in the main text, equation (B10). We stress again that each $\alpha$-element in the latter equations acts separate from the other ones at each time-interval we divided $\Delta t$. Neglecting in equation (B10) the counter-propagating waves and adopting a multi-rotating frame, similarly as in equation (A11) for the two-level case, we obtain:

$$H_I'^{(\text{red})}(t) \approx i(g\mu_b)^2 \sum_{\alpha,j\in\alpha} \frac{B_x b_{z,j}}{2\hbar\omega_{z,j}}\left[S_z^{(D)}, S_x\right]. \tag{B11}$$

## Appendix C. Suzuki–Trotter decomposition of a logical rotation

As discussed in the main text, the implementation of a Rabi oscillation requires to simultaneously address transitions between each eigenstate belonging to the $\ell = 0$ and to the $\ell = 1$ subspaces. In the illustrative $S = 3/2$ case reported in figure 3, $|0_L\rangle$ has support on $\{|1\rangle, |2\rangle\}$, while $|1_L\rangle$ on $\{|0\rangle, |3\rangle\}$.

It follows that the transition to be activated to perform a logical operation between $|0_L\rangle$ ans $|1_L\rangle$ are $\{(0,1),(0,2),(1,3),(2,3)\}$.

Due to the effects of the last term in (B6), if 4 pulses along $\hat{z}$ resonant with these transitions are provided, we also get oscillating drives at angular frequencies $\omega_{1,2} = \omega_{0,1} - \omega_{0,2}$ and $\omega_{0,3} = \omega_{0,1} + \omega_{1,3}$. These will induce an unwanted evolution between states $|0,0\rangle \leftrightarrow |0,1\rangle$ e $|1,0\rangle \leftrightarrow |1,1\rangle$, thus making the resulting transformation different from a logical gate and hence not fault-tolerant (FT).

This effect is avoided by separating the drive into two different part implemented in series via a Suzuki–Trotter decomposition. We just need to group the various transitions into groups such that each index appears exactly once: $\{(0,1),(2,3)\}$, $\{(0,2),(1,3)\}$.

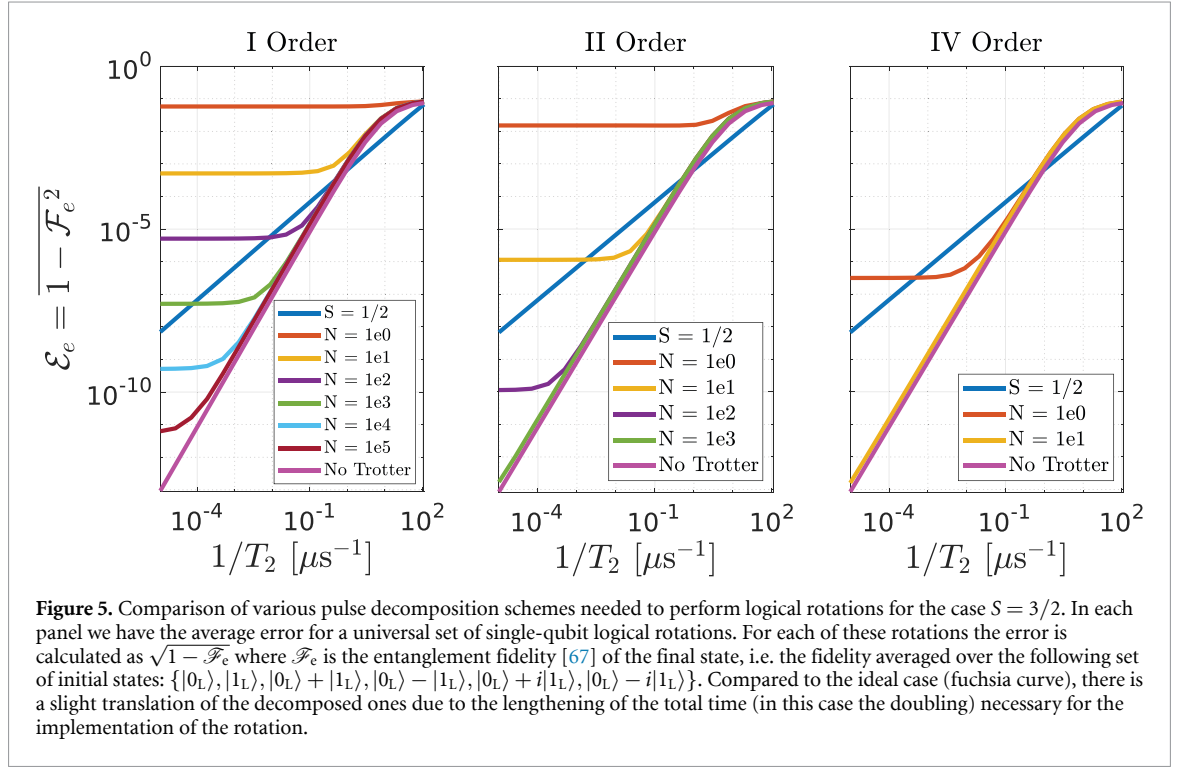

**Figure 5.** Comparison of various pulse decomposition schemes needed to perform logical rotations for the case $S = 3/2$. In each panel we have the average error for a universal set of single-qubit logical rotations. For each of these rotations the error is calculated as $\sqrt{1 - \mathcal{F}_e}$ where $\mathcal{F}_e$ is the entanglement fidelity [67] of the final state, i.e. the fidelity averaged over the following set of initial states: $\{|0_L\rangle, |1_L\rangle, |0_L\rangle + |1_L\rangle, |0_L\rangle - |1_L\rangle, |0_L\rangle + i|1_L\rangle, |0_L\rangle - i|1_L\rangle\}$. Compared to the ideal case (fuchsia curve), there is a slight translation of the decomposed ones due to the lengthening of the total time (in this case the doubling) necessary for the implementation of the rotation.

Figure 5 shows the average error associated with a universal set of single-qubit logical rotations for different Trotter schemes. This error is obtained by averaging the entanglement infidelity for each of the considered rotations. We recall that the definition of entanglement fidelity [67] is the average of the fidelity obtained by applying the quantum channel to the initial states $|0_L\rangle, |1_L\rangle, |0_L\rangle + |1_L\rangle, |0_L\rangle - |1_L\rangle, |0_L\rangle + i|1_L\rangle, |0_L\rangle - i|1_L\rangle$.

Regardless of the chosen scheme, the fault-tolerance of the procedure is preserved. Indeed, once the Trotter scheme is set and an adequate number of steps is taken, we can recover the performance of the ideal case. It should be noted, however, that set $\delta t$ as the time necessary to implement the ideal operation, the decomposed implementation would require a time equal to $n\delta t$. Note that $n$ is the number of terms into which the operator to be implemented is decomposed, and *not* the number of Trotter steps used. Therefore, this approach is ideal as it allows recovery of any desired precision for the implementation of the logical rotation simply by increasing the number of steps used.

These arguments can be directly extended to the case of a generic $S$. Specifically, given that $|0_L\rangle$ and $|1_L\rangle$ are defined over $\frac{2S+1}{2}$ disjoint energy eigenstates, the pulses required to perform a logical rotation must be divided into $\frac{2S+1}{2}$ groups. This ensures that no unwanted resonant terms are introduced by the term in equation (B6). Furthermore, since the number of terms in the decomposition scales linearly with $S$, the time required for a logical rotation will also increase linearly. As a result, the performance of the implemented code will deteriorate linearly with $S$. However, in the ideal case, as $S$ increases, there would be a quasi-exponential improvement in performance [18]. Therefore, even with the linear degradation in performance due to the Trotter decomposition, a quasi-exponential gain is still achieved as $S$ increases.

## Appendix D. FT quantum computing (QC)

We outline here how to implement FT-QC on the physical system considered here, referring for details to [18]. We consider a generic quantum operation $\mathscr{U}$ represented in the logical basis $\{|\ell, k\rangle\}$. This can be a logical gate (such as the logical Rabi used in the sensing protocol), the $\mathscr{S}$ for error detection or a recovery operator $\zeta_k$. In the case of a logical gate, this representation would be given by $U \otimes \mathbb{I}_{d/2}$.

First, we calculate the generator $\mathscr{G} = i \log \mathscr{U}$. This operation can be performed by a classical computer since the logical representation of $\mathscr{U}$ is simply a $d \times d$ matrix with $d = 2S + 1$. We then apply a basis change to move from the logical basis to the Hamiltonian eigenstates, represented by the matrix $\mathscr{A}$. The resulting generator on the physical system eigenstates is given by $\mathscr{A}\mathscr{G}\mathscr{A}^\dagger$.

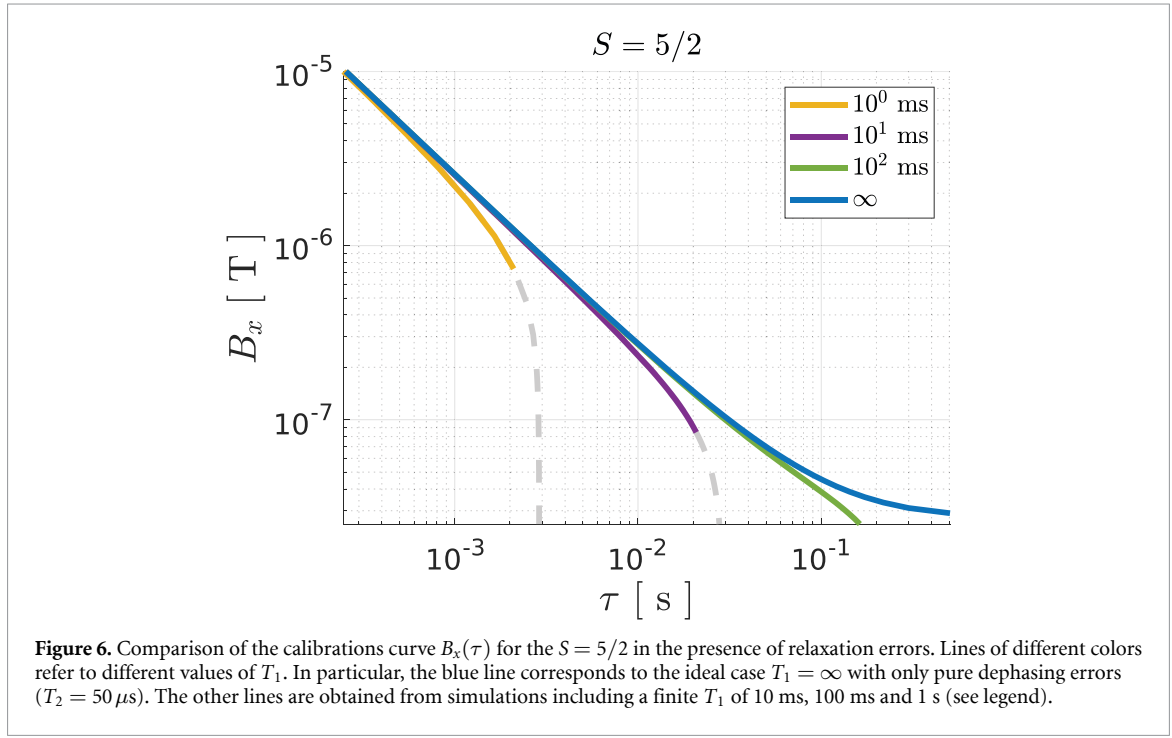

**Figure 6.** Comparison of the calibrations curve $B_x(\tau)$ for the $S = 5/2$ in the presence of relaxation errors. Lines of different colors refer to different values of $T_1$. In particular, the blue line corresponds to the ideal case $T_1 = \infty$ with only pure dephasing errors ($T_2 = 50\,\mu s$). The other lines are obtained from simulations including a finite $T_1$ of 10 ms, 100 ms and 1 s (see legend).

This generator allows us to assess the frequencies and amplitudes of the pulses required to achieve the desired evolution of the system. In the specific case considered here, due to the form of the eigenstates discussed in the main text, only resonant pulses along $\hat{z}$ are needed to execute the error detection and correction operations ($\mathscr{S}$ and $\zeta_k$). Conversely, the measured $B_x$ induces only logical rotations.

## Appendix E. Relaxation errors

Relaxation errors can be included in the simulated protocol by the following master equation (5) [68]

$$\dot{\rho}(t) = -\frac{i}{\hbar}[H, \rho(t)] + \sum_{a,b} \frac{W_{b \leftarrow a}}{T_1} \left( |b\rangle\langle a|\rho(t)|a\rangle\langle b| - \frac{1}{2}\{|a\rangle\langle a|, \rho(t)\} \right)$$
$$+ \sum_{a,b} \frac{\Gamma_{a,b}}{T_2} \left( |a\rangle\langle a|\rho(t)|b\rangle\langle b| - \frac{\delta_{a,b}}{2}\{|a\rangle\langle a|, \rho(t)\} \right). \tag{E1}$$

Here $\{,\}$ is the anti-commutator and $W$ is the rate matrix, obtained by considering transitions between system eigenstates $|a\rangle$ and $|b\rangle$ induced by modulations of the zero-field splitting Hamiltonian due to the coupling to the vibrational bath $W_{b \leftarrow a} = \alpha^2 |\langle b|O|a\rangle|^2$ (in the vanishing temperature limit). For simplicity, we have considered in our simulations a modulation of $O_2^1 = \{S_z, S_x\}$ extended Stevens operator, but results do not critically depend on that choice. In addition, we have assumed average spin-phonon coupling $\alpha^2$ which yields a decay of $|\Delta m| \approx 1$ on a timescale $T_1$. The last term of equation (E1) models pure dephasing. A detailed description of how to compute $\Gamma$ can be found in [18].

The resulting calibration curves $B_x(\tau)$ are shown in figure 6 for different values of $T_1$ for a $S = 5/2$ qudit. We note that spin relaxation errors do not hamper the protocol if they occur on long time scales compared to the interrogation time. This is typically the case for spin relaxation, which for molecular spin qubit/qudits can reach values in the order of 0.1–1 s at 5–10 K [49] and should become much longer below 1 K.

## Appendix F. Error operators and code words

In table 1 we report the diagonal Kraus operators modelling pure dephasing ($T_2$) obtained by process tomography of a free evolution of the qudit considered in the main text. In table 2 we report the codewords used in the simulations obtained by numerically solving the Knill–Laflamme conditions (7) with the Kraus operators of table 1.

**Table 1.** Kraus operators representing the pure dephasing channel for the physical system considered [18]. Process tomography of a free evolution of the system is used to derive the diagonal operators $E_k$. Each row lists the diagonal elements of $E_k$ in the basis of the eigenstates (columns) of the system.

| $S$ | $E_k$ | $|d_0\rangle\langle d_0|$ | $|d_1\rangle\langle d_1|$ | $|d_2\rangle\langle d_2|$ | $|d_3\rangle\langle d_3|$ | $|d_4\rangle\langle d_4|$ | $|d_5\rangle\langle d_5|$ | $|d_6\rangle\langle d_7|$ | $|d_7\rangle\langle d_7|$ |
|---|---|---|---|---|---|---|---|---|---|
| | | | | | Kraus operators | | | | |
| $\frac{3}{2}$ | $E_0$ | $9.9988 \times 10^{-1}$ | $9.9918 \times 10^{-1}$ | $9.9996 \times 10^{-1}$ | $9.9888 \times 10^{-1}$ | | | | |
| | $E_1$ | $1.5707 \times 10^{-2}$ | $4.0417 \times 10^{-2}$ | $-8.8440 \times 10^{-3}$ | $-4.7298 \times 10^{-2}$ | | | | |
| | $E_2$ | $-4.7091 \times 10^{-4}$ | $6.6400 \times 10^{-4}$ | $-7.4171 \times 10^{-4}$ | $5.4969 \times 10^{-4}$ | | | | |
| | $E_3$ | $8.2900 \times 10^{-6}$ | $-2.9673 \times 10^{-6}$ | $-6.8142 \times 10^{-6}$ | $1.4916 \times 10^{-6}$ | | | | |
| $\frac{5}{2}$ | $E_0$ | $9.9990 \times 10^{-1}$ | $9.9999 \times 10^{-1}$ | $9.9912 \times 10^{-1}$ | $9.9857 \times 10^{-1}$ | $9.9434 \times 10^{-1}$ | $9.9400 \times 10^{-1}$ | | |
| | $E_1$ | $1.3845 \times 10^{-2}$ | $6.8330 \times 10^{-4}$ | $4.2015 \times 10^{-2}$ | $-5.3461 \times 10^{-2}$ | $1.0607 \times 10^{-1}$ | $-1.0925 \times 10^{-1}$ | | |
| | $E_2$ | $-3.1148 \times 10^{-3}$ | $-3.3121 \times 10^{-3}$ | $-1.8661 \times 10^{-3}$ | $-1.5382 \times 10^{-3}$ | $5.2081 \times 10^{-3}$ | $4.6764 \times 10^{-3}$ | | |
| | $E_3$ | $3.9860 \times 10^{-5}$ | $-1.3702 \times 10^{-5}$ | $1.2853 \times 10^{-5}$ | $-1.7291 \times 10^{-4}$ | $-6.1284 \times 10^{-5}$ | $7.9509 \times 10^{-5}$ | | |
| | $E_4$ | $1.4722 \times 10^{-6}$ | $2.8763 \times 10^{-6}$ | $-3.1630 \times 10^{-6}$ | $-2.1725 \times 10^{-6}$ | $4.7478 \times 10^{-7}$ | $5.1222 \times 10^{-7}$ | | |
| | $E_5$ | $0$ | $0$ | $0$ | $0$ | $0$ | $0$ | | |
| $\frac{7}{2}$ | $E_0$ | $9.9818 \times 10^{-1}$ | $9.9845 \times 10^{-1}$ | $9.9404 \times 10^{-1}$ | $9.9424 \times 10^{-1}$ | $9.6577 \times 10^{-1}$ | $9.5100 \times 10^{-1}$ | $8.9567 \times 10^{-1}$ | $8.9287 \times 10^{-1}$ |
| | $E_1$ | $-2.9842 \times 10^{-2}$ | $-1.6304 \times 10^{-2}$ | $-9.9385 \times 10^{-2}$ | $9.5037 \times 10^{-2}$ | $-2.5911 \times 10^{-1}$ | $3.0871 \times 10^{-1}$ | $-4.3089 \times 10^{-1}$ | $4.4012 \times 10^{-1}$ |
| | $E_2$ | $5.2457 \times 10^{-2}$ | $5.3177 \times 10^{-2}$ | $4.4452 \times 10^{-2}$ | $4.8853 \times 10^{-3}$ | $-2.8714 \times 10^{-3}$ | $-1.3242 \times 10^{-2}$ | $-1.0968 \times 10^{-1}$ | $-9.4764 \times 10^{-2}$ |
| | $E_3$ | $8.8417 \times 10^{-4}$ | $-1.5394 \times 10^{-4}$ | $5.9143 \times 10^{-3}$ | $-8.2922 \times 10^{-3}$ | $1.1377 \times 10^{-2}$ | $-1.1130 \times 10^{-2}$ | $-8.2252 \times 10^{-3}$ | $9.6327 \times 10^{-3}$ |
| | $E_4$ | $7.2467 \times 10^{-4}$ | $7.7038 \times 10^{-4}$ | $2.2514 \times 10^{-4}$ | $4.7312 \times 10^{-4}$ | $-1.8018 \times 10^{-3}$ | $-1.7786 \times 10^{-3}$ | $5.9583 \times 10^{-4}$ | $7.9650 \times 10^{-4}$ |
| | $E_5$ | $-3.4355 \times 10^{-5}$ | $-1.5118 \times 10^{-5}$ | $-1.0196 \times 10^{-4}$ | $1.3951 \times 10^{-4}$ | $7.0422 \times 10^{-5}$ | $-6.3861 \times 10^{-5}$ | $-1.2559 \times 10^{-5}$ | $1.7919 \times 10^{-5}$ |
| | $E_6$ | $-3.4312 \times 10^{-6}$ | $-6.4727 \times 10^{-6}$ | $7.4387 \times 10^{-6}$ | $4.2274 \times 10^{-6}$ | $-1.3162 \times 10^{-6}$ | $-7.0161 \times 10^{-7}$ | $1.2323 \times 10^{-7}$ | $1.3242 \times 10^{-7}$ |
| | $E_7$ | $0$ | $0$ | $0$ | $0$ | $0$ | $0$ | $0$ | $0$ |

**Table 2.** For each values of S considered here are reported the encodings for codewords and errorwords of the QEC protocol used to correct for the pure dephasing of system considered [18]. Such encodings are obtained through a numerical optimization of the Knill–Laflamme conditions (7).

| | | | | Codewords and Errorwords | | | | | |
|---|---|---|---|---|---|---|---|---|---|
| S | $\lvert \ell, k \rangle$ | $\lvert d_0 \rangle$ | $\lvert d_1 \rangle$ | $\lvert d_2 \rangle$ | $\lvert d_3 \rangle$ | $\lvert d_4 \rangle$ | $\lvert d_5 \rangle$ | $\lvert d_6 \rangle$ | $\lvert d_7 \rangle$ |
| $\frac{3}{2}$ | $\lvert 0,0 \rangle$ | 0 | $8.5335 \times 10^{-1}$ | $5.2135 \times 10^{-1}$ | 0 | | | | |
| | $\lvert 0,1 \rangle$ | 0 | $5.2135 \times 10^{-1}$ | $-8.5335 \times 10^{-1}$ | 0 | | | | |
| | $\lvert 1,0 \rangle$ | $9.2188 \times 10^{-1}$ | 0 | 0 | $3.8748 \times 10^{-1}$ | | | | |
| | $\lvert 1,1 \rangle$ | $-3.8748 \times 10^{-1}$ | 0 | 0 | $9.2188 \times 10^{-1}$ | | | | |
| $\frac{5}{2}$ | $\lvert 0,0 \rangle$ | 0 | $8.6730 \times 10^{-1}$ | $4.8736 \times 10^{-1}$ | 0 | 0 | $1.0135 \times 10^{-1}$ | | |
| | $\lvert 0,1 \rangle$ | 0 | $-3.5270 \times 10^{-1}$ | $7.4532 \times 10^{-1}$ | 0 | 0 | $-5.6577 \times 10^{-1}$ | | |
| | $\lvert 0,2 \rangle$ | 0 | $-3.5127 \times 10^{-1}$ | $4.5495 \times 10^{-1}$ | 0 | 0 | $8.1831 \times 10^{-1}$ | | |
| | $\lvert 1,0 \rangle$ | $9.5163 \times 10^{-1}$ | 0 | 0 | $2.8743 \times 10^{-1}$ | $1.0852 \times 10^{-1}$ | 0 | | |
| | $\lvert 1,1 \rangle$ | $1.9924 \times 10^{-1}$ | 0 | 0 | $-8.4622 \times 10^{-1}$ | $4.9417 \times 10^{-1}$ | 0 | | |
| | $\lvert 1,2 \rangle$ | $-2.3388 \times 10^{-1}$ | 0 | 0 | $4.4865 \times 10^{-1}$ | $8.6256 \times 10^{-1}$ | 0 | | |
| $\frac{7}{2}$ | $\lvert 0,0 \rangle$ | 0 | $9.3589 \times 10^{-1}$ | $3.4838 \times 10^{-1}$ | 0 | 0 | $4.9608 \times 10^{-2}$ | $1.6842 \times 10^{-2}$ | 0 |
| | $\lvert 0,1 \rangle$ | 0 | $2.6564 \times 10^{-1}$ | $-7.7686 \times 10^{-1}$ | 0 | 0 | $5.2233 \times 10^{-1}$ | $-2.3043 \times 10^{-1}$ | 0 |
| | $\lvert 0,2 \rangle$ | 0 | $1.6583 \times 10^{-1}$ | $-3.0968 \times 10^{-1}$ | 0 | 0 | $-7.7601 \times 10^{-1}$ | $-5.2384 \times 10^{-1}$ | 0 |
| | $\lvert 0,3 \rangle$ | 0 | $-1.6139 \times 10^{-1}$ | $4.2335 \times 10^{-1}$ | 0 | 0 | $3.5002 \times 10^{-1}$ | $-8.1989 \times 10^{-1}$ | 0 |
| | $\lvert 1,0 \rangle$ | $9.7463 \times 10^{-1}$ | 0 | 0 | $2.0861 \times 10^{-1}$ | $7.9230 \times 10^{-2}$ | 0 | 0 | $1.7359 \times 10^{-2}$ |
| | $\lvert 1,1 \rangle$ | $-1.2072 \times 10^{-1}$ | 0 | 0 | $7.6080 \times 10^{-1}$ | $-5.7741 \times 10^{-1}$ | 0 | 0 | $2.7057 \times 10^{-1}$ |
| | $\lvert 1,2 \rangle$ | $1.3493 \times 10^{-1}$ | 0 | 0 | $-3.0685 \times 10^{-1}$ | $-7.1823 \times 10^{-1}$ | 0 | 0 | $-6.0974 \times 10^{-1}$ |
| | $\lvert 1,3 \rangle$ | $1.3160 \times 10^{-1}$ | 0 | 0 | $-5.3246 \times 10^{-1}$ | $-3.8008 \times 10^{-1}$ | 0 | 0 | $7.4479 \times 10^{-1}$ |

## ORCID iDs

Matteo Mezzadri  https://orcid.org/0000-0002-2800-2695
Luca Lepori  https://orcid.org/0000-0003-2323-4988
Alessandro Chiesa  https://orcid.org/0000-0003-2955-3998
Stefano Carretta  https://orcid.org/0000-0002-2536-1326